\documentclass[preprint,amsmath,amssymb,aps]{revtex4}

\usepackage{slashed}
\usepackage{amsmath}
\usepackage{amsthm}
\usepackage{mathtools}
\usepackage[utf8]{inputenc}
\usepackage{graphicx}
\usepackage{epsfig}
\usepackage{amssymb}
\usepackage{dcolumn}
\usepackage{bm}
\usepackage{color}
\usepackage[dvipsnames]{xcolor}

\newcommand{\ba}{\begin{array}}
\newcommand{\ea}{\end{array}}
\def\br{\begin{eqnarray}}
\def\er{\end{eqnarray}}
\def\be{\begin{equation}}
\def\ee{\end{equation}}
\usepackage{hyperref}

\def\({\left(}
\def\){\right)}

\def\<{\left\langle}
\def\>{\right\rangle}

\begin{document}

\title{A natural QCD infrared cutoff \footnote{To appear in the Festschrift for Ruben Aldrovandi} }

\author{A. A. Natale} 
\email{adriano.natale@unesp.br}

\affiliation{Instituto de F{\'i}sica Te\'orica - UNESP, Rua Dr. Bento T. Ferraz, 271,\\ Bloco II, 01140-070, S\~ao Paulo, SP, Brazil}

\begin{abstract}
We briefly discuss some results obtained recently about dynamical gluon mass generation. We comment that this mass provides a natural QCD infrared cutoff 
and also implies an infrared finite coupling constant. We also discuss the phenomenological applications of these results and how they 
can be treated in the context of the so-called Dynamical Perturbation Theory.

\end{abstract}

\maketitle



\section{ Introduction: Dynamical gluon mass generation}

In recent years, following a conjecture proposed forty years ago \cite{cor1}, we have seen a great development in the understanding of the Schwinger mechanism in QCD \cite{ag1,bo1,ag2,bi1,pa1,ag3}.
In quantum electrodynamics in two space-time dimensions it is relatively simple to verify the presence of the Schwinger mechanism, that is, the presence of a pole in the scalar part of the polarisation of the dimensionless vacuum. In the QCD case such verification is much more sophisticated. This means that a particular sum of diagrams involving propagators and vertices ends up resulting in a dynamical gluon mass. This result obtained through solutions 
of the Schwinger-Dyson equations (SDE) is strongly confirmed when compared with results obtained from lattice QCD \cite{ag1}. The observation of a dynamical gluon mass implies the study of the gluon self-energy, that is, the vacuum
polarisation, which involves the three-gluon vertex, which in turn has a longitudinally coupled part with a simple pole structure
\be
{\Gamma}^{pole}_{\alpha\mu\nu} (q,r,p) \propto \frac{q_\alpha}{{q}^{2}} \delta_{\mu\nu} C_1(q,r,p)+...
\label{eq1}
\ee
where $C_1(q,r,p)_{{q}^{2}\approx{0}}  = 2q.r {\mathbb{C}}({r}^{2})+ O({q}^{2})$.

This pole structure can be related to the scalar function ${\mathbb{C}}(r^2)$ that can be extracted from lattice QCD data. 
Without going into details, and directing the reader directly to Ref.\cite{ag3}, this scalar function must be no null if the Schwinger mechanism is operative 
in QCD, and this was demonstrated with extraordinary accuracy in Ref.\cite{ag3}, as shown in Fig.(1)

\begin{figure}[h]
\centering
\includegraphics[scale=0.5]{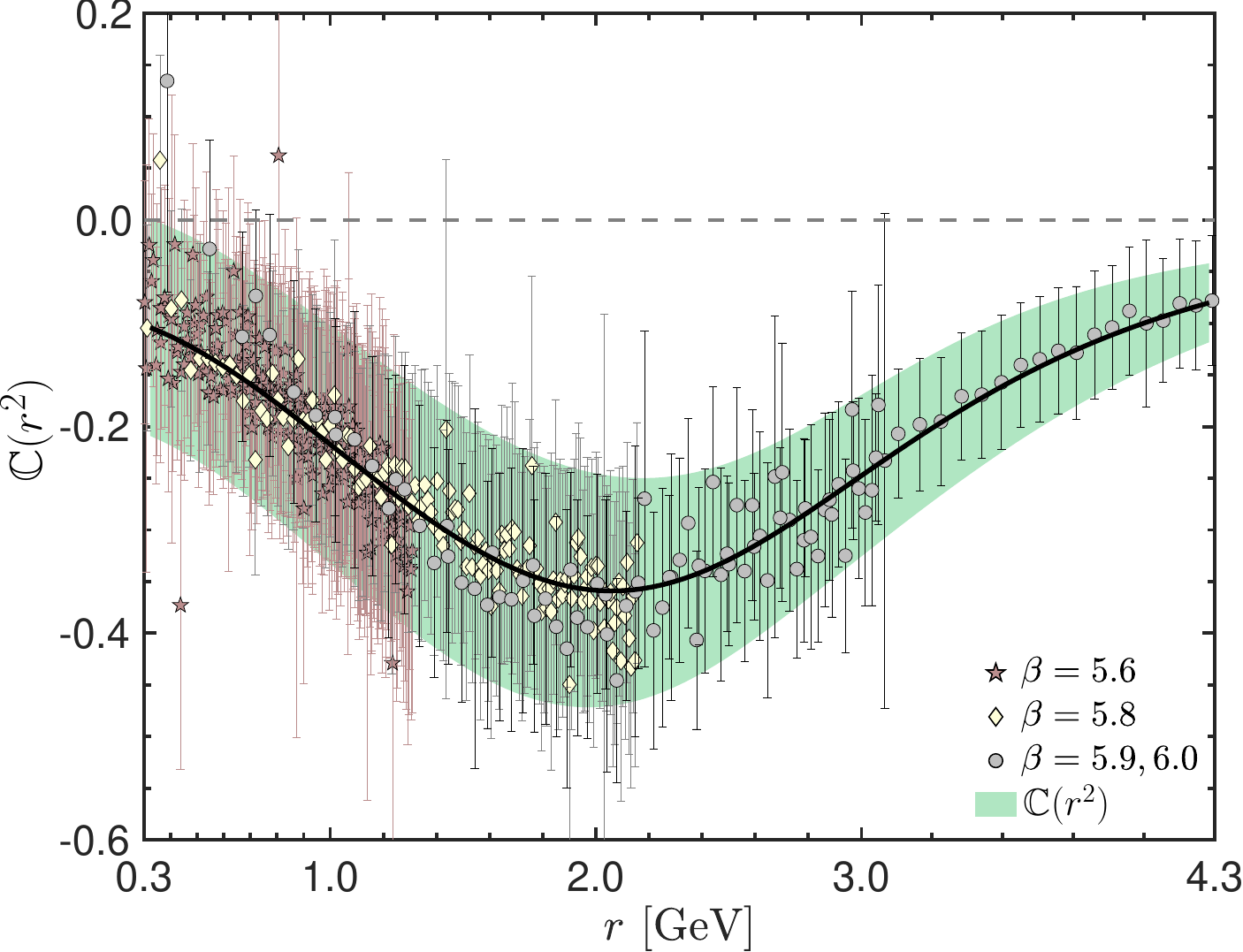}
\vspace{-0.25cm} 
\caption[dummy0]{ ${\mathbb{C}}(r^2)$ obtained in Ref.\cite{ag3}. Solid black curve - obtained using central fit
forms of lattice QCD of Ref.\cite{ag4}. The
bracketing soft-green band expresses the uncertainty in this result. Dashed grey line - null result (no Schwinger mechanism).}
\label{fig1}
\end{figure}

The former analysis that has been done to determine the existence of a dynamical gluon mass \cite{ag00} is becoming more and more complex, and it is difficult to imagine that these results, which indicate a natural infrared cutoff in QCD, are going to be easily 
taken into account in phenomenological applications. However, models that assume massive gluons to understand the infrared QCD behavior began to emerge (see, for instance, Ref.\cite{ma1,ma2}),
and they are indicating that if we take into account the existence of this natural cutoff, several infrared QCD calculations are well behaved when treated with perturbative methods \cite{ma2,ma3}. 
The possibility of an expansion in small parameters as advocated in Ref. \cite{ma2,ma3} is not entirely surprising. The phenomenon of dynamical gluon mass generation is also associated with an infrared 
fixed point of the coupling constant ($\alpha_s$) \cite{aa1}, and, although the strong force is proportional to the product of $\alpha_s$ times the gluon propagator, the isolated value of the coupling constant is important \textit{per se}, and there are phenomenological estimates 
that the infrared value of the coupling constant may not be so large, as will be discussed ahead. The important fact is that these QCD complex and detailed studies
come out with an infrared finite gluon propagator and an infrared finite coupling constant.

\section{ Phenomenology with dynamically massive gluons}

It is clear that in phenomenological calculations at large energies quarks and gluons are treated as free particles, and usually the latter enters the calculation as a massless particle. Although the perturbative QCD 
analysis in that energy range works quite well, it would be very interesting if we had a method to treat QCD where both its non-perturbative and perturbative 
aspects were covered, and a fair amount of work has been done in this 
direction \cite{rob1,rob2}. Indeed in some phenomenological calculations a gluon mass 
must be invoked if we are to describe the experimental data. As one example we can recall the hadronic decay of heavy quarkonium like $V=\Upsilon, \, J/\psi $ that can be measured through the branching ratio
$R_V = {\Gamma(V\rightarrow ggg)}/{\Gamma(V\rightarrow ee)}$, which can also provide a measurement of the QCD coupling constant. If the gluon has a dynamical mass 
(whose value at small momentum will be
indicated by $m_g$) the branching ratio $R_V$ must be changed to $A = R_V \cdot f_3(\eta)$ where 
\be
f_3(\eta)=\frac{\Gamma(V\rightarrow ggg)_{m_g}}{\Gamma(V \rightarrow ggg)_{m_g=0}},
\label{eq11}
\ee
with $\eta=2m_g/M_V$ and  $M_V$ is the quarkonium mass. The function $f_3(\eta)$ decreases the value of $A$ and is a measure of the effect of the dynamical gluon mass as discussed in Ref.\cite{mi1}. This 
effect is even more remarkable in the data of the radiative process $V \rightarrow \gamma + X$, where QCD predicts a photon spectrum nearly linear in $z=2E_\gamma/M_V$ ($E_\gamma$ is the 
photon energy). In this case, the photon spectrum does not reach the maximum value of $z$, indicating a suppression due to the fact that the phase space is occupied by gluons that behave like massive 
particles \cite{par1,co1}.

Another example where the dynamical gluon mass plays a fundamental role is in the calculation of hadronic cross sections at high energies in the soft regime,
which are dominated by Pomeron exchange. The simplest Pomeron construction in QCD is given by a two-gluon exchange which shows a singularity at $-t=0$.
To solve this problem Landshoff and Nachtmann (LN) suggested that the gluon propagator is intrinsically modified in the infrared region \cite{la1}.
The LN model was improved with the introduction of dynamically massive gluons in Ref.\cite{an1,an2}, producing a good description of the elastic differential 
cross section for $pp$ scattering data at $\sqrt{s}=53$ GeV. It is clear that the Pomeron component to this cross section should be even more dominant
in the LHC regime, where other Reggeon contributions are negligible. A reanalysis of the LN Pomeron model at TeV energies, taking into account the results of dynamical 
gluon mass generation was performed recently \cite{bop1}, where the scattering amplitude is giving by
\begin{eqnarray}
 {\cal A}(s,t) = is^{\alpha_{\Bbb P}(t)} \frac{1}{\tilde{s}_{0}}\frac{8}{9}n_{p}^{2} [ \tilde{T}_{1} - \tilde{T}_{2} ] ,
\label{ampliyyy}
\end{eqnarray}
with
\begin{widetext}
\begin{eqnarray}
\tilde{T}_{1} = \int_{0}^{s} d^{2}k \, \bar{\alpha}\left( \frac{q}{2} + k \right) D\left( \frac{q}{2} + k \right) \bar{\alpha}\left( \frac{q}{2} - k \right) D\left( \frac{q}{2} - k \right) \left[ G_{p}(q,0)  \right]^{2} ,
\label{t1}
\end{eqnarray}
\begin{eqnarray}
\tilde{T}_{2} = \int_{0}^{s} d^{2}k \, \bar{\alpha}\left( \frac{q}{2} + k \right) D\left( \frac{q}{2} + k \right) \bar{\alpha}\left( \frac{q}{2} - k \right) D\left( \frac{q}{2} - k \right) G_{p}\left( q,k - \frac{q}{2}\right)  \times \nonumber \\
\hspace*{0.5cm} \left[ 2 G_{p}(q,0) - G_{p}\left( q,k - \frac{q}{2}\right)  \right] .
\label{t2}
\end{eqnarray}
\end{widetext}
Here $\alpha_{\Bbb P}(t)$ is the Pomeron trajectory, $\tilde{s}_{0} \equiv s_{0}^{\alpha_{\Bbb P}(t)-1}$ , $G_{p}(q,k)$ is a convolution of proton wave functions,
and $n_{p}=3$ is the number of quarks in the proton.  $T_{1}$ ($T_{2}$) represent the contribution when both gluons attach to the same quark (to different quarks) within the proton, and the coupling $\bar{\alpha}(q^2)$ and gluon propagator $D(q^2)$ are the infrared finite ones that have been obtained in the study of dynamical mass generation that we referred above. In Figs.(2) and (3) we present the fits for the elastic scattering data obtained in Ref.\cite{bop1} for the LHC experiments ATLAS and TOTEM, where we use different fits of the dynamical gluon masses obtained in the literature, which are indicated by $m_{log}(q^{2})$ and $m_{pl}(q^{2})$, and
verify that the data is well described by this Pomeron model. The result is dependent on the infrared value of the gluon mass, but still not precise enough to determine the cross section dependence on the formal expression of the dynamical gluon mass with momentum. 

\begin{figure}\label{fig001}
\begin{center}
\includegraphics[height=.35\textheight]{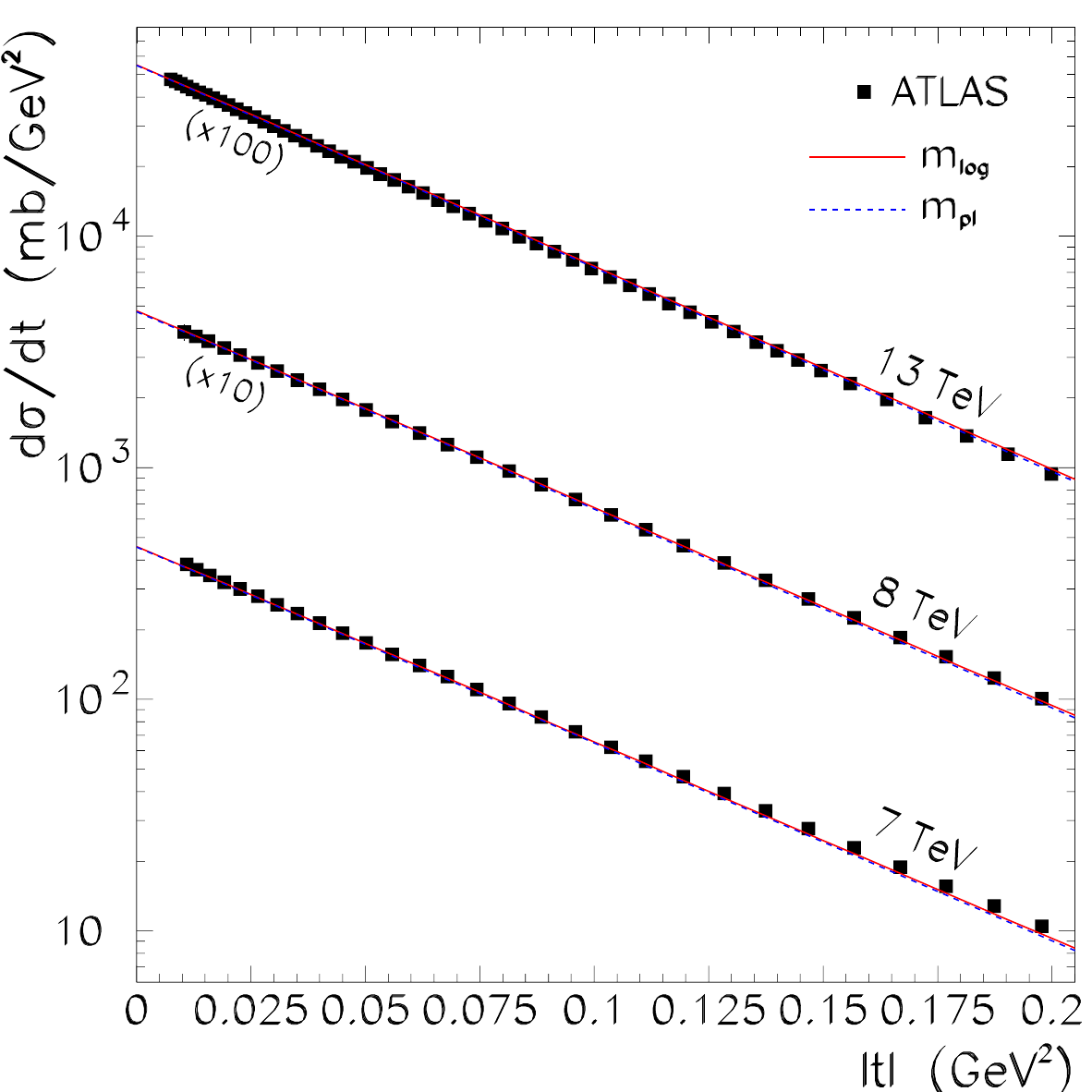}
\caption{Pomeron model description of the $pp$ elastic differential cross section data from ATLAS. The solid and dashed lines show the results obtained 
using fits of the dynamical gluon mass $m_{log}(q^{2})$ and $m_{pl}(q^{2})$, respectively.}
\end{center}
\end{figure}

\begin{figure}\label{fig002}
\begin{center}
\includegraphics[height=.35\textheight]{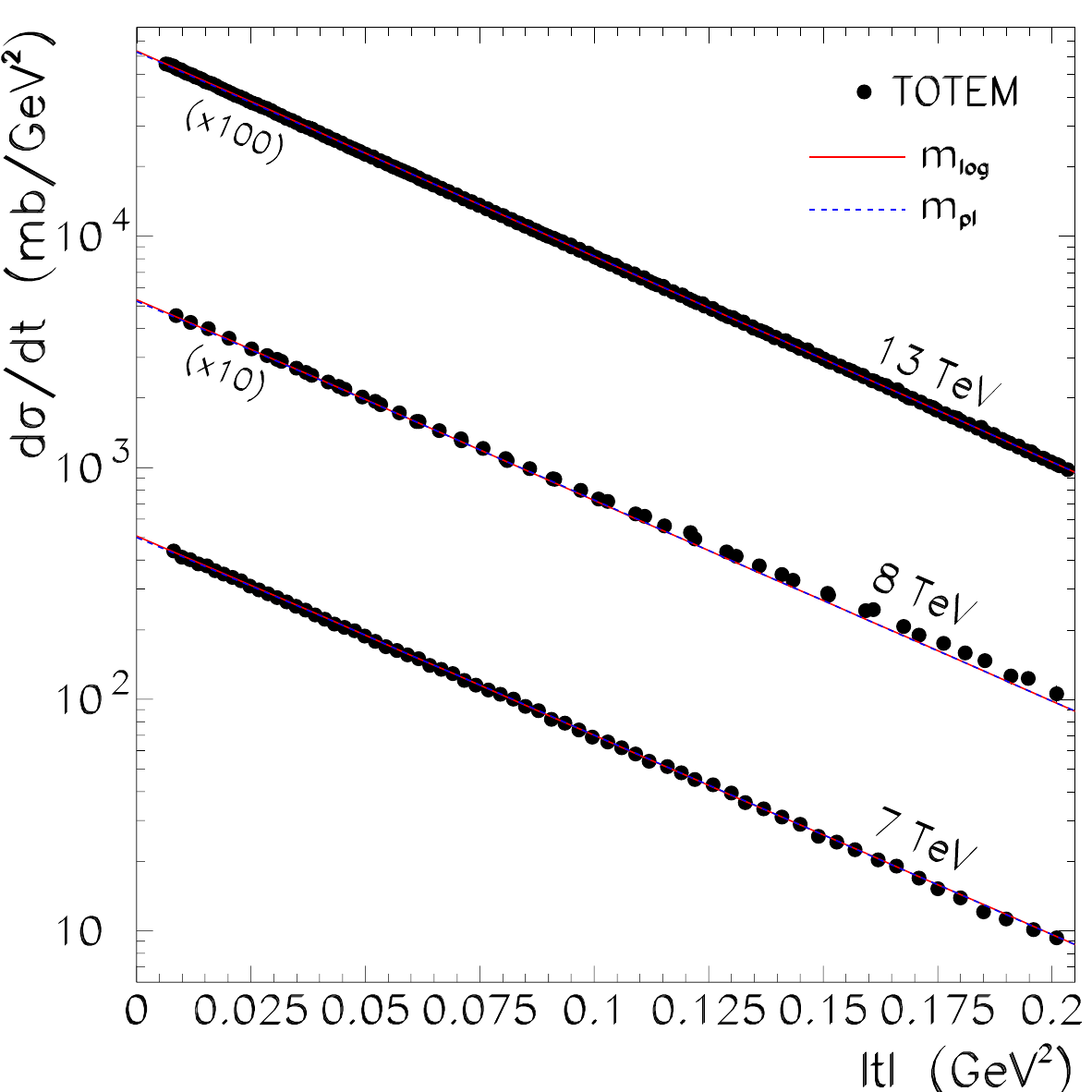}
\caption{Pomeron model description of the $pp$ elastic differential cross section data from TOTEM. The solid and dashed lines show the results obtained 
using fits of the dynamical gluon mass $m_{log}(q^{2})$ and $m_{pl}(q^{2})$, respectively.}
\end{center}
\end{figure}

Looking at the examples we quoted above and others \cite{ag5}, we can say that the consequences of introducing the phenomenon of dynamical mass generation for the gluon cannot be totally neglected in phenomenological calculations. We can say even more: the consideration of these effects in loop calculations (where we have integrations over the whole range of momenta) may even eliminate the existence of the so-called renormalons \cite{ben1}.

\section{ Dynamical perturbation theory}

The perturbative description of gauge theory quantities in terms of a coupling constant is of
central importance to our understanding of such theories, and thereby of our ability to use
them for phenomenology. Much of the current high-energy experimental data needs to be compared with theoretical 
phenomenological calculations involving various orders of perturbation theory. This means that they can involve QCD calculations 
at the level of several loops, mixing the non-perturbative part with the perturbative one. The direct calculation of such
non-perturbative corrections is in many cases very challenging, and methods like the ones used to compute
the dynamical gluon mass (see, for instance, Refs.\cite{ag2,pa1}) have not been introduced into standard phenomenological calculations,
and due to its complexity it is possible that it will take a long time to be incorporated into the high energy hadronic phenomenology.

It is interesting that a great interest in non-perturbative effects in quantum field theory has recently arisen in what is described as the 
resurgence method \cite{ec1}, although this technique is still far from being applied in realistic high energy phenomenological 
models \cite{re1,re2,re3}. Therefore, a simpler method of obtaining phenomenological results, using the fact that we have a natural infrared QCD cutoff 
(the dynamical gluon mass) and an infrared finite strong coupling constant would be very welcome. One proposal in that direction
was formulated many years ago by Pagels and Stokar \cite{pag1}, and was denominated by Dynamical Perturbation Theory (DPT).

DPT is a generalization of perturbation theory, and as stated in Ref.\cite{pag1} it can be described as follows. Amplitudes that do not vanish in
all orders of perturbation theory are given by their free field values. Amplitudes that vanish in all orders in perturbation theory
like $\lambda = e^{-1/bg^2}$, as, for example, the dynamical gluon mass that decays with momentum $p$ like $1/p^2$, are retained in the series,
in comparison with higher orders like $g^n e^{-1/bg^2}$ with $n>0$, which can be less important due to the soft behavior of the dressed amplitudes.
The use of a dressed gluon propagator and coupling constant, as obtained in solutions of SDE, in the calculations described
in Section II is exactly an application of DPT. Note that DPT still assumes a perturbative expansion, and, in favor of this approximation, we can say that there are estimates that the coupling constant $g$ may have a moderate value \cite{co2,ag7,go3}, as well as the fact that models assuming massive gluons admit a well-behaved expansion when compared with results obtained in lattice QCD \cite{ma2,ma3,si1,si2}, but it is clear that more studies on this possibility are needed.

\section{Final remarks}

The existence of the Schwinger mechanism in QCD is strongly proven by comparing the results obtained through SDE
and those obtained in lattice QCD calculations \cite{ag1,ag3}. The numerical SDE solutions indicating dynamical gluon mass generation are
basically performed in the Landau gauge, this solution in other gauges has been discussed in Refs.\cite{cor1,ag2,ag6}. 

The Landau gauge makes the SDE calculations much simpler, remembering that this type of calculation is still quite laborious to be taken into account in phenomenological calculations. 
At the present time, the practical way to use this information is to use simple fits for the various Green functions, that have been obtained in
SDE solutions associated with the Schwinger mechanism, following the DPT proposal.
The use of approximate functions in phenomenological calculations is interesting because different phenomenological data depend 
differently on propagators and vertices. Therefore, the set of experimental data can also serve as a test to determine the functional behavior of the dynamical
gluon mass, as well as the coupling constant, whose infrared value is dependent on $m_g$. The examples of hadronic phenomenology presented in Section II show this different dependence of physical quantities as a function of the gluon propagator and QCD coupling constant.

There are other examples of high energy phenomenology that can be modified when considering the existence of a dynamical gluon mass and infrared finite coupling constant \cite{ex1,ex2,ex3}. The presence of an infrared fixed point in QCD \cite{aa1} can modify the evolution of the coupling constants in the study of grand 
unification models \cite{go1}, and it can modify the determination of the conformal region in technicolor or similar theories with many fermions \cite{go2}. 
The effect of dynamically massive gluons also causes different effects on the chiral transition of quarks in the fundamental and adjoint representations \cite{cap1}, 
and also affects the determination of structure functions at small-x \cite{lu1}.

In this quite brief report we emphasized the existence of the Schwinger mechanism in QCD, how it can modify different observables of hadronic phenomenology, and that the use of the so-called Dynamic Perturbation Theory may be the simplest method to utilize the results of dynamical gluon mass generation.

\section*{Acknowledgments}

I have benefited from discussions with A. C. Aguilar, E. G. S. Luna and A. Doff.

\end{document}